\documentclass[aps,pra,showpacs,twocolumn]{revtex4}
\usepackage{graphicx}
\usepackage{bm}
\usepackage{amssymb}
\usepackage{amsmath}
\usepackage{times}

\newcommand{\md}{\,\mathrm{d}} 

\newcommand{\trace}{\mathop{\rm Tr}\nolimits}

\newcommand{\pr}{\mbox{Pr}} 

\newcommand{\cL}{{\mathcal L}} 
\newcommand{\cN}{{\mathcal N}} 

\newcommand{\E}{{\mathbb{E}}}

\newcommand{\Pe}{{\mathbb{P}}}
\DeclareRobustCommand\openone{\leavevmode\hbox{\small1\normalsize\kern-.33em1}}
\newcommand{\id}{\mathrm{\openone}}

\newcommand{\be}{\begin{equation}}
\newcommand{\ee}{\end{equation}}
\newcommand{\bea}{\begin{eqnarray}}
\newcommand{\eea}{\end{eqnarray}}
\newcommand{\beas}{\begin{eqnarray*}}
\newcommand{\eeas}{\end{eqnarray*}}

\newcount\minute
\newcount\hour
\def\currenttime{%
    \minute\time
    \hour\minute
    \divide\hour60
    \the\hour:\multiply\hour60\advance\minute-\hour\the\minute}

\begin{document}
\title{Statistical Inference from Imperfect Photon Detection}
\author{Koenraad M.R.~Audenaert}
\email{koenraad.audenaert@rhul.ac.uk}
\affiliation{Dept.\ of Mathematics, Royal Holloway, University of London,
Egham, Surrey TW20 0EX, UK}
\author{Stefan Scheel}
\email{s.scheel@imperial.ac.uk}
\affiliation{Quantum Optics and Laser Science, Blackett Laboratory,
Imperial College London, Prince Consort Road, London SW7 2AZ, UK}
\begin{abstract}
We consider the statistical properties of photon detection with
imperfect detectors that exhibit dark counts and less than unit
efficiency, in the context of tomographic reconstruction. In this
context, the detectors are used to implement certain POVMs that would
allow to reconstruct the quantum state or quantum process under
consideration. Here we look at the intermediate step of inferring
outcome probabilities from measured outcome frequencies, and show how
this inference can be performed in a statistically sound way in the
presence of detector imperfections. Merging outcome probabilities for
different sets of POVMs into a consistent quantum state
picture has been treated elsewhere [K.M.R.~Audenaert and S.~Scheel,
New J.\ Phys.\ \textbf{11}, 023028 (2009)]. Single-photon pulsed
measurements as well as continuous wave measurements are covered.
\end{abstract}

\date{\today, \currenttime}

\pacs{03.67.-a,42.50.-p,42.50.Ct}
\maketitle

\section{Introduction}
Estimating quantum states and processes plays an increasingly
important role in quantum engineering as it allows for an unambiguous
verification of the generation and manipulation procedures applied to
a quantum system. Amongst the plethora of reconstruction methods, only
few are capable of specifying error bars associated with the
reconstruction process itself. We have recently developed a Kalman
filtering approach to quantum tomographic reconstruction
\cite{kalman1} based on Bayesian analysis employing a linear Gaussian
noise model.

In Ref.~\cite{kalman1} we have dealt with quantum state and process
reconstruction from tomographic data obtained by perfect measurements.
In optical tomography, for example, this corresponds to the assumption
that detectors are perfect and detector counts represent photon counts
faithfully. In reality, however, optical detectors are not perfect and
exhibit dark counts and losses (less than unit efficiency). In
addition, mode mismatch in the detector connection may lead to further
losses.

In the context of tomographic reconstruction these imperfections have
important consequences. The detectors form part of an implementation
of a POVM $\{\Pi^{(1)},\Pi^{(2)},\ldots,\Pi^{(k)}\}$, with which one
endeavours to estimate, say, a quantum state $\rho$. The measurements
consist of frequencies $\bm{g}=(g_1,g_2,\ldots,g_K)$ for each outcome
$i=1,2,\ldots,K$. Each of these frequencies corresponds to a
probability $p_i=\trace\rho\Pi^{(i)}$. To estimate $\rho$, one
essentially first estimates the probabilities
$\bm{p}=(p_1,p_2,\ldots,p_K)$ from the frequencies $\bm{g}$.
In the context of Bayesian inference, the estimation procedure yields
a probability distribution for $\bm{p}$ given the measured
$\bm{g}$. Indeed, only in the limit of an infinite number $N$ of
measurements do the relative frequencies $\bm{g}/N$ tend to the
probabilities $\bm{p}$. For finite $N$, $\bm{p}$ cannot be known with
perfect certainty, and, hence, must be described as a random variable
with a certain distribution. Bayesian inference tells us what this
distribution should be.

In Ref.~\cite{kalman1} we have shown how knowledge of this
distribution can ultimately lead to a reconstruction of the state in
terms of a probability distribution over state space; one thus obtains
a confidence region, rather than a single point in state space, as in
maximum-likelihood methods. The basic tool for this reconstruction is
the Kalman filter equation. It requires as input the first and second
moments of the distribution of $\bm{p}$ inferred from $\bm{g}$, for
the various POVMs used in the tomography.

Detector imperfections are important in this respect because they have
an impact on the inferred distribution of $\bm{p}$. The measurement
mean $\bm{z}$ taken in by the Kalman filter has to reflect losses and
dark counts. Equally important is that imperfections lead to
additional measurement fluctuations which have to be accounted for in
the measurement covariance matrix $\bm{\Theta}$.

In this article we present a statistically sound method for
incorporating detector imperfections in the reconstruction scheme. One
of the main design goals is practicality, and speed, without
sacrificing statistical accuracy too much. In particular, we want to
avoid lengthy numerical calculations at all costs, excluding any
method that reeks of Monte Carlo. 
To this purpose we aim at finding
exact formulas for the required quantities, or if that is impossible,
we introduce several approximation methods to reduce the computational
complexity of finding the quantities numerically.

The article is organised as follows. We present a
mathematical model for an imperfect detector in
Sec.~\ref{sec:model}. In Sec.~\ref{sec:pulses} we treat the first case
of optical detectors used in a setup where the optical beam consists
of timed single-photon pulses. The continuous wave setup is treated in
Sec.~\ref{sec:poisson}. We also study, in Sec.~\ref{sec:param}, how
one can incorporate imprecisions in the parameters that describe the
detector imperfections, dark count rate and efficiency. We conclude
with a brief overview of the main results obtained, in
Sec.~\ref{sec:conclusion}. Appendix \ref{app:A} is devoted to a  numerical
method for calculating certain integrals that are needed for the
calculation of the moments of the distribution of $\bm{p}$.
In  appendix \ref{sec:prelim}, we
gather the necessary definitions for a number of special functions and
special distributions that are used extensively in the paper. 
A number of implementation notes are also given. 
\section{Modelling the photon detection process\label{sec:model}}

In this section we present a physical model for an imperfect photon
detector and review how the statistical properties of such a detector
comes about, for further reference. We assume throughout that the
detector operates in Geiger mode, so that photon detection consists of
single-detection events, as opposed to linear mode where an
opto-electrical current is produced.

\begin{figure}[ht]
\includegraphics[width=6cm]{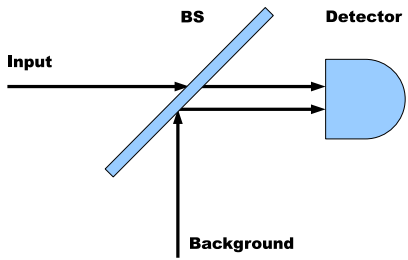}
\caption{
\label{fig:bs}
Model of an imperfect detector.
}
\end{figure}
In the theory of quantum detection, an imperfect detector exhibiting
dark counts is modeled by a compound detector, consisting of a perfect
detector set in one of the outgoing arms of a beam splitter. This beam
splitter mixes incoming light fields with a background radiation field
(see Fig.~\ref{fig:bs}). The efficiency $\eta$ of the actual detector
is modeled by the transmission coefficient $|T|^2$ of the beam
splitter. The background radiation field is assumed to be coupled to a
thermal bath, and is best described as a multi-mode field.

Under the additional and well-justified assumption that the number of
modes in the background field is much larger than the number of photons,
coupling between background modes and incoming modes can be ignored
(see, e.g. Refs.~\cite{semenov} and \cite{mandel} p.~681). Under this
assumption, the background photon distribution is approximately
Poissonian. We will assume that the mean value of the number of dark
counts per measurement interval is known, a value denoted by
$\alpha$. Thus, the number of dark counts per measurement interval is
a random variable $R\sim \Pe(\alpha)$.

Given this physical model, the detection statistics can be derived as
follows. The conditional probability that the detector produces
$m$ counts given that $n$ photons are present in the incoming field
and $r$ photons in the background field
is given by \cite{hong}
\bea
f_{M|N,R}(m|n,r) &=& f_{M|N,R}(m-r|n,0) \nonumber\\
&=& {n\choose m-r}\eta^{m-r}(1-\eta)^{n-m+r},
\eea
where the binomial coefficient is taken to be 0 whenever $r>m$ or
$m-r>n$. Since under the given assumption the background photon
distribution is approximately Poissonian, we set $R\sim \Pe(\alpha)$
and obtain
\be
f_{M|N}(m|n) = e^{-\alpha} \sum_{r=0}^m \frac{1}{r!} {n\choose m-r}
\alpha^r\eta^{m-r}(1-\eta)^{n-m+r}.
\ee

For a given photon number distribution of the incoming light field,
$f_N(n)$, the distribution of the photon counts is
\be
f_M(m) = \sum_{n=0}^\infty f_{M|N}(m|n)\,f_N(n).
\ee
One verifies easily that if the incoming light field is Poissonian,
$N\sim\Pe(\nu)$, with $f_N(n)=\exp(-\nu)\nu^n/n!$, the distribution of
$M$ is Poissonian also, $M\sim\Pe(\alpha+\eta\nu)$, as expected.

If the incoming light field is in a Fock state, with either $n=0$ or
$n=1$, the formulas reduce to
\be
f_{M|N}(m,0) = e^{-\alpha}\frac{\alpha^m}{m!}
\ee
for $n=0$ (no input photon), and
\be
f_{M|N}(m,1) = e^{-\alpha}\left[(1-\eta)\frac{\alpha^m}{m!} +\eta
\frac{\alpha^{m-1}}{(m-1)!}\right]
\ee
for $n=1$ (single input photon). With short laser pulses, one usually
only wants to discriminate between $m=0$ and $m\neq 0$ (let alone that
further discrimination is at all possible). Hence one is only
interested in
\be
f_{M|N}(0,0) = e^{-\alpha},
\ee
(no click, no input photon), and
\be
f_{M|N}(0,1) = e^{-\alpha}(1-\eta),
\ee
(no click, 1 input photon), and their complementary values. Usually,
$\alpha$ is rather small, and one can set $e^{-\alpha}\approx 1-\alpha$.

\section{Single Photon Pulses\label{sec:pulses}}

In this section we treat the case of a pulsed laser beam, where each
pulse consists of a single photon. The statistics of the detection
events are governed by the binomial or multinomial distribution. We
treat three different setups. First, a 2-outcome POVM where only one
detector is used; the second detector, for the second outcome, is left
out on the assumption that the total number of detection events should
be equal to the number of pulses anyway. For perfect detectors, this
assumption is correct, while in the presence of detector imperfections
this is only an approximation. We will study how this affects the
detection statistics. Next, we treat a 2-outcome POVM with both
detectors in use and compare it with the previous case. Finally, a
$K$-outcome POVM is considered, generalising the $K=2$ case.
\subsection{Single Detector\label{sec:singleA}}
\begin{figure}[ht]
\includegraphics[width=6cm]{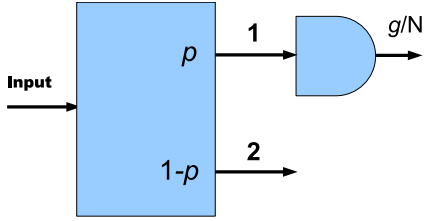}
\caption{
\label{fig:det1}
Model of a 2-outcome POVM where only one detector is used.
}
\end{figure}

We first consider the most simple case of a 2-outcome POVM where only
one detector is used.
The tomographic apparatus, apart from the detectors, is hereby treated
as a black box with $2$ output terminals, one for each POVM element,
and we assume that in each of the $N$ runs, for a fixed setting of the
POVM, a single photon appears at one of the output terminals. Losses
in the tomographic apparatus itself are disregarded, because that is
inessential for the derivation of the detector model. The tomography
black box can thus be modeled by a $2$-dimensional probability
distribution $\bm{p}=(p,1-p)$, where $p$ represents the probability
that the photon appears at terminal $1$ (see Fig.~\ref{fig:det1}).

Terminal 1 is then connected to a detector with dark count rate
$\alpha$ and efficiency $\eta$, while terminal 2 is left open; this
corresponds to the cheapest implementation of a 2-outcome
detector. The record of an $N$-run experiment consists of the number
of times $g$ the detector has clicked.

\subsubsection{Statistical model}
We first derive the statistical properties of the random variable $G$,
whose observations are the recorded photon count $g$. Its distribution
is conditional on $P$ and depends on the parameters $\alpha$ and
$\eta$.
The standard procedure is to first derive the conditional
probabilities of a detector clicking or not clicking conditional on a
photon coming in or not. These are given by (cf. Sec.~\ref{sec:model}):
\bea
\pr(1|0) &=& \pr(\text{click}|\text{no photon})
= \alpha, \nonumber \\
\pr(0|0) &=& \pr(\text{no click}|\text{no photon})
= 1-\alpha, \nonumber \\
\pr(0|1) &=& \pr(\text{no click}|\text{photon})
= \beta, \nonumber \\
\pr(1|1) &=& \pr(\text{click}|\text{photon})
= 1-\beta. \label{eq:singlecond}
\eea
Here we have introduced the \textit{attenuation factor} $\beta$ as
\be
\beta := (1-\alpha)(1-\eta).
\ee

Using these conditional probabilities we can calculate the
probability $q$ that the detector clicks:
\beas
q &:=& \pr(\text{click}) \\
&=& \pr(\text{click}|\text{no photon}) \pr(\text{no photon})\\
&& \mbox{}+\pr(\text{click}|\text{photon}) \pr(\text{photon}) \\
&=& \alpha(1-p) +(1-\beta)p \\
&=& \alpha+(1-\alpha-\beta)p \\
&=& \alpha+\gamma p,
\eeas
where in the last line we defined $\gamma$ as the slope of the $q$
versus $p$ curve, $\gamma:=1-\alpha-\beta = (1-\alpha)\eta$.

From this probability, one directly obtains the probability that in
$N$ runs $g$ clicks are counted given the probability $p$ of an
incoming photon. Obviously, $g$ should be an integer between 0 and
$N$. The conditional probability distribution of the count $G$,
conditional on $P$, is just the binomial distribution
$\text{Bin}(N;q)$ with probability distribution function (PDF)
\be
f_{G|P}(g|p) = {N \choose g}\,q^{g}(1-q)^{N-g}.
\label{eq:Pg1}
\ee

\subsubsection{Statistical Inference}
From the general formula (\ref{eq:Pg1}) describing the statistical
behaviour of an imperfect detector we can derive the likelihood
function $L_{P|G}$ that is needed for the Bayesian inference
procedure. It is immediately clear from Eq.~(\ref{eq:Pg1}) that the
likelihood function of $\alpha+(1-\alpha-\beta) P$ will be
proportional to the PDF of a beta-distribution with parameters $a=g+1$
and $b=N-g+1$. To that we can add some prior information: $P$ is
restricted to the interval $[0,1]$. This implies that the
beta-distribution of $\alpha+\gamma P$ will have to be truncated to
the interval $[\alpha,1-\beta]$.

The moments of this truncated beta-distribution are given by
(with $\E[X]$ denoting the expectation value of a random variable $X$)
\bea
m_1 &:=& \E[\alpha+\gamma P]   \nonumber\\
&=&
\frac{B(\alpha,1-\beta,g+2,N-g+1)}{B(\alpha,1-\beta,g+1,N-g+1)}, \\
m_2 &:=& \E[(\alpha+\gamma P)^2] \nonumber\\
&=& \frac{B(\alpha,1-\beta,g+3,N-g+1)}{B(\alpha,1-\beta,g+1,N-g+1)}.
\eea
Here, $B(x_0,x_1,a,b)$ is the generalised incomplete beta function.
In actual numerical computations, it is better to use the
\textit{regularised} incomplete beta function $I_{x_0,x_1}(a,b)$.
Exploiting the relation $B(a+1,b)/B(a,b) = a/(a+b)$, we then get
\bea
m_1 &=& \frac{I_{\alpha,1-\beta}(g+2,N-g+1)}{I_{\alpha,1-\beta}(g+1,N-g+1)}
\,\, m_{1,0}, \label{eq:m1}\\
m_2 &=& \frac{I_{\alpha,1-\beta}(g+3,N-g+1)}{I_{\alpha,1-\beta}(g+1,N-g+1)}
\,\,m_{2,0}, \label{eq:m2} \\
m_{1,0} &=& \frac{g+1}{N+2}, \\
m_{2,0} &=& \frac{(g+1)(g+2)}{(N+2)(N+3)},
\eea
where the first factor in Eqs.~(\ref{eq:m1}) and (\ref{eq:m2})
is a correction term that goes to 1 when $\alpha$ and $\beta$ tend to
0, that is, for ideal detectors. From these expressions, the central
moments of $P$ can then be calculated as
\bea
\mu(P|G=g) &=& \frac{m_1-\alpha}{1-\alpha-\beta}, \label{eq:mup1} \\
\sigma^2(P|G=g) &=& \frac{m_2 -m_1^2}{(1-\alpha-\beta)^2}.
\label{eq:sigp1}
\eea

\begin{figure}[ht]
\includegraphics[width=8cm]{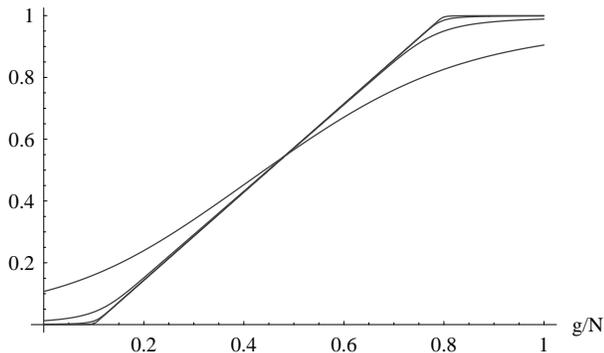}
\caption{
\label{fig:1}
Plot of $\mu(P|G=g)$ [Eq.~(\ref{eq:mup1})] as a function of $g/N$ for
$N=10$, $100$, $1000$, and $10000$, and values of $\alpha=0.1$ and
$\beta=0.2$.
}
\end{figure}

Figure~\ref{fig:1} shows a plot of $\mu$ [Eq.~(\ref{eq:mup1})] as a
function of $g/N$ for a few values of $N$. As could be expected, for
sufficiently large $N$, the curve for $\mu$ approaches a piecewise
linear curve with $\mu=0$ for $0\le g/N\le \alpha$, and $\mu=1$ for
$1-\beta\le g/N\le 1$.

\begin{figure}[ht]
\includegraphics[width=8cm]{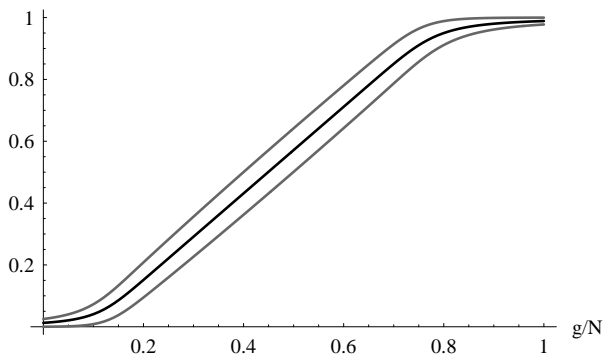}
\caption{
\label{fig:2}
Plot of $\mu(P|G=g)$ [Eq.~(\ref{eq:mup1})] (black, central curve)
and $\sigma(P|G=g)$ [Eq.~(\ref{eq:sigp1})] (depicted as the grey curves
$\mu\pm\sigma$) as a function of $g/N$ for $N=100$
and values of $\alpha=0.1$ and $\beta=0.2$.
}
\end{figure}

Figure \ref{fig:2} singles out the case $N=100$ and depicts the values
of the first and second central moments, $\mu$ and $\sigma$.

\subsubsection{Discussion}
A common way to deal with dark counts and non-unit detector efficiency
is to subtract the dark count rate $\alpha$ from the relative count
frequencies $g/N$, replacing negative numbers by 0 if necessary, and
then divide by $1-\alpha-\beta$, replacing numbers higher than 1 by 1,
if necessary. In other words, one would use formula (\ref{eq:mup1})
with $g/N$ in place of $m_1$, and truncate the outcome to the interval
$[0,1]$.

We argue that there are two distinct problems with this approach.
First, as we have already argued in Ref.~\cite{kalman1}, for given
$g$, the inferred distribution of $P$ is a beta (Dirichlet)
distribution, not a binomial (multinomial) distribution. Considering
the extremal case $g/N\le\alpha$, the above method would assign 0 to
the probability $P$, which amounts to claiming that the outcome can
never happen (except for dark counts). Of course, never having seen an
event does not imply that the event is impossible. Indeed, the correct
approach, using the beta distribution, assigns non-zero mean and
variance to $P$. Second, as can be seen from Figs.~\ref{fig:1} and
\ref{fig:2}, the actual behaviour of the statistically correct
inferences for $P$ vary smoothly with $g$ and the truncation mentioned
above is only correct in the $N\to \infty$ limit.

\subsection{A $2$-outcome experiment with $2$ detectors\label{sec:singleB}}
\begin{figure}[ht]
\includegraphics[width=6cm]{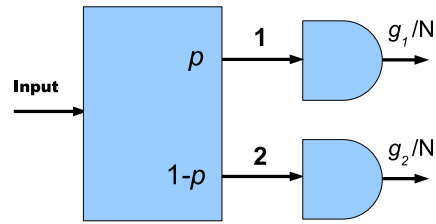}
\caption{
\label{fig:det2}
Model of a 2-outcome POVM where 2 detectors are used.
}
\end{figure}

In this Section, we consider the situation where a single photon can
take one of two paths (with probability $p$ and $1-p$, respectively),
and subsequently impinges on one of two detectors, each set along one
path (Fig.~\ref{fig:det2}).

\subsubsection{Statistical Model}
Let detector $i$ be characterised by a dark rate $\alpha_i$ and an
attenuation factor $\beta_i$. Concerning the presence of the photon at
the detectors, there are two exclusive events: event $10$, where the
photon is at detector $1$ and not at detector $2$, or event $01$,
where the photon is at detector 2 instead. Concerning the detectors
clicking, there are 4 events: $00$, $10$, $01$ and $11$, corresponding
to no detector clicking, only detector 1 clicks, only detector 2
clicks, or both detectors are clicking. We stress again that we are
considering single-photon experiments, hence the latter case of both
detectors clicking would typically correspond to one detector
detecting the photon just mentioned while the other detector is
producing a dark count. With perfect detectors such an event would not
occur.

The corresponding conditional probabilities are easily calculated.
Let $\pr(ij|kl)$ denote this conditional probability, where $i=1$
iff detector 1 clicks, $j=1$ iff detector 2 clicks, $k=1$ iff a photon
is at detector 1, and $l=1$ iff a photon is at detector 2; hence,
$k+l=1$. Because the two detectors are independent, we have
$\pr(ij|kl)=\pr(i|k)\,\pr(j|l)$, where $\pr(\cdot|\cdot)$ is the
single-detector conditional probability (\ref{eq:singlecond}) of the
previous section.

Combined with the probability of the photon events $10$ and $01$ being
$\pr(k=1)=p$ and $\pr(l=1)=1-p$, this gives the probabilities of the
click events:
\bea
q_{00} &=& p \beta_1(1-\alpha_2) + (1-p)(1-\alpha_1)\beta_2, \\
q_{01} &=& p \beta_1\alpha_2 +(1-p)(1-\alpha_1)(1-\beta_2), \\
q_{10} &=& p(1-\beta_1)(1-\alpha_2)+(1-p)\alpha_1\beta_2, \\
q_{11} &=& p(1-\beta_1)\alpha_2 +(1-p)\alpha_1(1-\beta_2).
\eea
The probabilities of the corresponding event frequencies $g_{00}$,
$g_{01}$, $g_{10}$ and $g_{11}$, counting over $N$ runs, is given by
the multinomial distribution
\be
f_{\bm{G}|P} = {N \choose g_{00},g_{01},g_{10},g_{11}}
q_{00}^{g_{00}}
q_{01}^{g_{01}}
q_{10}^{g_{10}}
q_{11}^{g_{11}}.
\ee
Note that if one does not distinguish between single click events and
two-click events, one is capturing the sums $g_{01}+g_{11}$ and
$g_{10}+g_{11}$, in which the 2-click events are counted twice. This
causes mathematical difficulties in the statistical inference process
that are best avoided.

One may actually discard the multiple-click events altogether, and
only record the single-click events $g_1:=g_{10}$ and
$g_2:=g_{01}$. This means that one makes no distinction between
$g_{00}$ and $g_{11}$. The corresponding distribution is again
multinomial, but now given by
\be
f_{G_1,G_2|P} =
{N \choose g_1,g_2,g_0}
q_{10}^{g_1}
q_{01}^{g_2}
(q_{00}+q_{11})^{g_0}.
\ee
with $g_0=N-g_1-g_2$.

In the special case that both detectors are identical, i.e.\ when they
have the same
dark count rates and attenuation factors,
$\alpha_1=\alpha_2=\alpha$, and
$\beta_1=\beta_2=\beta$, we find that the third factor $q_{00}+q_{11}$
reduces to the constant $\beta(1-\alpha)+\alpha(1-\beta)$, independent
of $p$. Then, considered as a function of $p$, $f_{G_1,G_2|P}$ is
proportional to the binomial PDF ${g_1+g_2\choose
g_1}q_{10}^{g_1}q_{01}^{g_2}$, with
\bea
q_{10} &=& (1-p)\alpha\beta+p(1-\alpha)(1-\beta), \\
q_{01} &=& p \alpha\beta +(1-p)(1-\alpha)(1-\beta).
\eea
Defining
\bea
a_1&:=&\alpha\beta,\\
a_2&:=&(1-\alpha)(1-\beta),
\eea
we have
$q_{10}=a_1+(a_2-a_1)p$ and
$q_{01}=a_2-(a_2-a_1)p$.
Furthermore, by defining
\be
a:=\frac{a_1}{a_1+a_2},
\ee
we find that
$q_{10}=(a_1+a_2)[a+(1-2a)p]$ and
$q_{01}=(a_1+a_2)[a+(1-2a)(1-p)]$.

Thus, the PDF $f_{G_1,G_2|P}$ is proportional to the truncated
binomial PDF:
\be
f_{G_1,G_2|P} \propto {g_1+g_2 \choose g_1}
[a+(1-2a)p]^{g_1}[a+(1-2a)(1-p)]^{g_2}.
\label{eq:f2equal}
\ee
This PDF is essentially identical to the PDF (\ref{eq:Pg1})
obtained in the previous section, apart from the fact that the dark
count rate $\alpha$ and the attenuation factor $\beta$ only enter in
the PDF via the single constant $a$. This constant assumes the role of
an \textit{effective dark count rate} and is given by
\be
a=\frac{a_1}{a_1+a_2} = \frac{\alpha\beta}{(1-\alpha)(1-\beta)+\alpha\beta}.
\label{eq:a}
\ee
One sees that $a$ is of the order of $\alpha\beta$, which is a smaller
number than $\alpha$ and $\beta$. More precisely, we have
$\alpha\beta\le a\le 2\alpha\beta$.

\subsubsection{Statistical Inference}
In general, the statistical inference formulas become quite
complicated, because in the expression for $f_{G_1,G_2|P}$ more than 2
factors appear that have a dependence on $p$. The subsequent integrals
over $p$ can no longer be expressed as (incomplete) beta functions. In
this section we treat the easiest case of all detectors being equal,
and use the PDF (\ref{eq:f2equal}), which only has two factors. As
this PDF is essentially identical to the PDF (\ref{eq:Pg1}) obtained
in the previous section, the same results therefore hold for the
statistical inference.

We can therefore use formulas (\ref{eq:m1})--(\ref{eq:sigp1}),
provided we perform the substitutions $g\to g_1$, $N\to g_1+g_2$,
$\alpha\to a$ and $\beta\to a$. This gives
\bea
m_1(a) &=& \frac{I_{a,1-a}(g_1+2,g_2+1)}{I_{a,1-a}(g_1+1,g_2+1)}
\,\,m_{1,0},\label{eq:m1a}\\
m_2(a) &=& \frac{I_{a,1-a}(g_1+3,g_2+1)}{I_{a,1-a}(g_1+1,g_2+1)}
\,\,m_{2,0},\\
m_{1,0} &=& \frac{g_1+1}{g_1+g_2+2}, \\
m_{2,0} &=& \frac{(g_1+1)(g_1+2)}{(g_1+g_2+2)(g_1+g_2+3)}
\eea
and
\bea
\mu(P|G=(g_1,g_2)) &=& \frac{m_1-a}{1-2a},\\
\sigma^2(P|G=(g_1,g_2)) &=& \frac{m_2 - m_1^2}{(1-2a)^2}.
\label{eq:sigp2}
\eea

The main difference between Eqs.~(\ref{eq:m1a})--(\ref{eq:sigp2}) and
Eqs.~(\ref{eq:m1})--(\ref{eq:sigp1}) for the single-detector case is
the replacement of $\alpha$ and $1-\beta$ as limits of the incomplete
beta functions by $a$ and $1-a$, where $a$ is the effective dark count
rate given by Eq.~(\ref{eq:a}).
\subsubsection{Discussion}
\begin{figure}[ht]
\includegraphics[width=8cm]{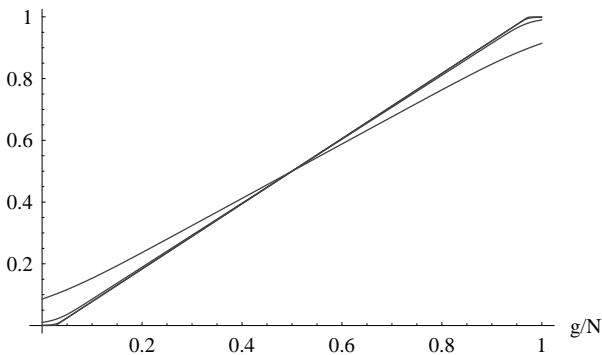}
\caption{
\label{fig:3}
Plot of $\mu(P|G=(g,N-g))$ as a function of $g/N$ for $N=10$, $100$,
$1000$, and $10000$, and values of $\alpha=0.1$ and $\beta=0.2$.
}
\end{figure}

\begin{figure}[ht]
\includegraphics[width=8cm]{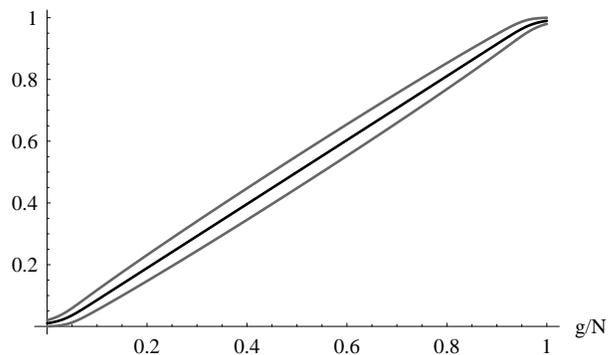}
\caption{
\label{fig:4}
Plot of $\mu(P|G=(g,N-g))$ and $\sigma(P|G=(g,N-g))$ (depicted as the
grey curves $\mu\pm\sigma$) as a function of $g/N$ for $N=100$ and
values of $\alpha=0.1$ and $\beta=0.2$.
}
\end{figure}

We can compare the performance of the two setups, one detector or two
detectors, by comparing the average value of $\sigma$ of the
reconstructed distribution of $p$, for a given value of the actual
$p$. In the 1-detector case, $G$ is distributed according to
Eq.~(\ref{eq:Pg1}). For given actual $p$, one calculates the average
of $\sigma$ as given by Eq.~(\ref{eq:sigp1}) over this
distribution. In the 2-detector case, $G_1,G_2$ are distributed
according to Eq.~(\ref{eq:f2equal}), and one similarly calculates the
average of $\sigma$ as given by Eq.~(\ref{eq:sigp2}). Taking, as in
Fig.~\ref{fig:1}, $\alpha=0.1$, $\beta=0.2$ and $N=100$, we find, for
various settings of the actual $p$, the values collected in
Tab.~\ref{tab:1}.
%
\begin{table}[h]
\begin{tabular}{l|cc}
$p$ & $\sigma$(1-detector) & $\sigma$(2-detectors) \\ \hline
0 & 0.033 & 0.044 \\
0.5 & 0.070 & 0.088 \\
1 & 0.04 & 0.044
\end{tabular}
\caption{\label{tab:1} Average values for $\sigma$ for various
settings of $p$, comparing the 1-detector and 2-detector cases.}
\end{table}
%
%
It emerges that the one-detector case performs slightly better on
average. Presumably, this is because for the 2-detector case we only
used single click events to keep the inference procedure simple. That
is, the sum $g_1+g_2$ is always less than $N$. With the given
parameter settings, the average value of $g_1+g_2$ is 50 (for any
$p$). However, if one makes more measurement runs for the 2-detector
case, stopping when $g_1+g_2$ is equal to the number of runs for the
1-detector case, the 2-detector setup performs better (Tab.~\ref{tab:2}).
%
\begin{table}[h]
\begin{tabular}{l|cc}
$p$ & $\sigma$(1-detector) & $\sigma$(2-detectors) \\ \hline
0 & 0.033 & 0.017 \\
0.5 & 0.070 & 0.052 \\
1 & 0.04 & 0.017
\end{tabular}
\caption{\label{tab:2} Average values for $\sigma$ in the 1-detector
and 2-detector protocols under the additional constraint that
$g_1+g_2$ equals the number of runs for the 1-detector case.}
\end{table}
%
%
In Figs.~\ref{fig:3} and \ref{fig:4} we show what happens to
Figs.~\ref{fig:1} and \ref{fig:2} for the 2-detector setup (under the
constraint $g_1+g_2=100$). The plateaus around $\mu=0$ and $\mu=1$ are
indeed much shorter. In addition, the error bars (quantified by
$\sigma$) are smaller by a factor of roughly $1/\sqrt{2}$
(corresponding to an on average increase of $N$ by a factor of 2).

\subsubsection{Unequal Detectors}
In the more realistic case that detector parameters are not equal, we
need to calculate integrals of the form
$$
J(\bm{g};\bm{a},\bm{b})=\int_0^1 dp\,\prod_i (a_i+b_i p)^{g_i},
$$
with more than 2 factors. Indeed, the mean of $P$ can be calculated from
$$
J_1= \E[a_1+b_1P] =
\frac{J(\bm{g}+\bm{e}^1;\bm{a},\bm{b})}{J(\bm{g};\bm{a},\bm{b})},
$$
and its variance from
$$
J_2=\E[(a_1+b_1P)^2] =
\frac{J(\bm{g}+2\bm{e}^1;\bm{a},\bm{b})}{J(\bm{g};\bm{a},\bm{b})},
$$
where $\bm{e}^i$ denotes the unit vector along the $i$-th dimension,
$\bm{e}^i=(0,\ldots,1,\ldots,0)$. Hence,
\beas
\mu_P&=&(J_1-a_1)/b_1, \\
\sigma^2_P&=&(J_2-a_1^2-2a_1b_1\mu_P)/b_1^2-\mu_P^2.
\eeas
The actual integrations can be performed numerically using standard
quadrature methods (e.g.\ Matlab's built-in {\tt quadl} routine).

To enhance numerical robustness for higher values of $N=\sum_i g_i$,
for which the integrand is sharply peaked, it is advisable to reduce
the integration interval and only integrate over that subinterval of
$[0,1]$ where the integrand is higher than, say, $10^{-6}$ times its
maximal value. This refinement allows the quadrature algorithm to
better place its quadrature points.

\subsection{A $K$-outcome POVM with $K$ Detectors\label{sec:mtnK}}
Here, we generalise the results of Sec.~\ref{sec:singleB} to the
case where there are $K$ detectors, each one corresponding to one of
the outcomes. No detector is missing. The tomographic apparatus is now
treated as a black box with $K$ output terminals, one for each POVM
element. To keep the calculations for the statistical inference
transparent, we restrict ourselves to the case of identical detectors
throughout.

\subsubsection{Statistical Model}
Again we assume that in each of the $N$ runs, for a fixed setting of
the POVM, a single photon appears at one of the output terminals.
The tomography black box is now modeled by a $K$-dimensional
probability distribution $\bm{p}=(p_k)_{k=1}^K$, where $p_k$
represents the probability that the photon appears at terminal $k$.

Each terminal is then connected to a detector with dark count rate
$\alpha$, efficiency $\eta$, and attenuation factor $\beta$. The
record of an $N$-run experiment consists of the frequencies $g_k$,
$k=1,\ldots,K$, the number of times the $k$-th detector has clicked
and none of the others has. As discussed before, we leave out events
where more than one detector clicked, in order not to increase the
mathematical complexity.

We now derive the statistical properties of the vector
$\bm{G}=(G_k)_{k=1}^K$, whose observations are the recorded photon
counts $\bm{g}$. Its distribution is conditional on the probability
vector $\bm{p}$ and depends on the parameters $\alpha$ and $\beta$.

Let $q_k$ denote the probability of the event $E_k$ that detector $k$
clicks and no other. We again first calculate the conditional
probabilities of $E_k$, conditional on the photon appearing at
terminal $j$. For $j=k$, this conditional probability is
$(1-\alpha)^{K-1}(1-\beta)$; for $j\neq k$ it is
$\alpha\beta(1-\alpha)^{K-2}$.

The probability of event $E_k$ is then given by
\bea
q_k &=& (1-\alpha)^{K-2}[(1-\alpha)(1-\beta) p_k + \alpha\beta(1-p_k)]
\nonumber \\
&=& \frac{a_1+(a_2-a_1) p_k}{(K-1)a_1+a_2} \nonumber \\
&=& a+(1-Ka)p_k,
\eea
where $a_1$ and $a_2$ are defined as before, and the effective dark
count rate $a$ is defined as
\be
a := \frac{a_1}{(K-1)a_1+a_2}.
\ee

For the $N$-run experiment, the probability of the vector of
frequencies $\bm{g}=(g_1,g_2,\ldots,g_K)$ is therefore proportional to
the truncated multinomial distribution
\be
f_{\bm{G}|\bm{P}}(\bm{g}|\bm{p})
\propto {\sum_k g_k \choose g_1,\ldots,g_K} \prod_{i=1}^K
[a+(1-Ka) p_i]^{g_i}.
\label{eq:fgpmulti}
\ee
\subsubsection{Statistical Inference}
From the general formula (\ref{eq:fgpmulti}) describing the
statistical behaviour of a bank of imperfect detectors we immediately
derive that the likelihood function $L_{\bm{P}|\bm{G}}$ is given by
\be
L_{\bm{P}|\bm{G}} = \frac{1}{\cN}\,\prod_{i=1}^K [a+(1-Ka) p_i]^{g_i},
\ee
where $\cN$ is the normalisation integral, given by the integral of
$[a+(1-Ka) p_1]^{g_1} \ldots [a+(1-Ka) p_K]^{g_K}$ over the
probability simplex $p_k\ge0$, $\sum_k p_k=1$. This integral is quite
hard to calculate, and so are the integrals that are required to
calculate the moments of $L_{\bm{P}|\bm{G}}$. Denoting
\be
r_k := a+(1-Ka) p_k,\label{eq:rfromp}
\ee
we get that the random vector $\bm{R}=(R_1,\ldots,R_K)$ is distributed
according to a \textit{truncated Dirichlet distribution}, where $R_k$
is subject to the condition $R_k\ge a$.

No analytic expression is known for the integrals involved; among the
numerical methods to calculate them are numerical integration, the
Gibbs sampling method (a Monte Carlo method) \cite{boyer07}, and
saddle-point approximations \cite{walrus}. Since for neither method
commonly available software seems to exist, we give some more details
about the latter method in Appendix \ref{app:A}, where we calculate the
normalisation integral of the truncated Dirichlet distribution
$$
J(\bm{\alpha};a) := P(\bm{R}\ge a) =
\int_{\genfrac{}{}{0pt}{}{r_i\ge a}{\sum_i r_i=1}}
d\bm{r} f_{\bm{R}}(\bm{r}),
$$
for $\bm{R}\sim \mbox{Dir}(\bm{\alpha})$, where as usual
$\bm{\alpha}=\bm{g}+1$. The first and second order moments about the
origin of $\bm{R}$ can be expressed in terms of this integral as
\bea
\E[R_i] &=& \frac{J(\bm{\alpha}+\bm{e}^i;a)}{J(\bm{\alpha};a)}\,
\frac{\alpha_i}{\alpha_0} \\
\E[R_i^2] &=& \frac{J(\bm{\alpha}+2\bm{e}^i;a)}{J(\bm{\alpha};a)}\,
\frac{\alpha_i(\alpha_i+1)}{\alpha_0(\alpha_0+1)}\\
\E[R_i R_j] &=&
\frac{J(\bm{\alpha}+\bm{e}^i+\bm{e}^j;a)}{J(\bm{\alpha};a)}\,
\frac{\alpha_i\alpha_j}{\alpha_0(\alpha_0+1)}.
\eea
The moments of $\bm{P}$ then follow easily from Eq.~(\ref{eq:rfromp}).

Note, however, that this calculation requires $K+1+K(K+1)/2$ separate
integrations, which can be computationally very expensive for larger
values of $K$. For relatively small values of $a$, say $a<0.1$, the
following provides a moderately good approximation:
\be
J(\bm{\alpha};a)\approx \prod_{i=1}^K
I_{1-a}(\alpha_0-\alpha_i,\alpha_i)
\label{eq:Jbybeta}
\ee
with $I_{1-a}(\alpha_0-\alpha_i,\alpha_i)$ the regularised incomplete
beta function [see Eq.~(\ref{eq:regincbeta})].
Numerical experiments indicate that this approximation is good enough
for the calculation of the second order moments of $\bm{R}$ for values
of $a$ as large as $0.1$. This has been checked for $K=3$; with
$a=0.1$, the approximated second order moment differs less than 5\%
from its actual value.
Similarly, the first order moments are accurate to within $0.1\sigma$
for $a\le0.05$.

The worst case figures appear for extremal values of $\bm{G}$, i.e.\
all $g_i=0$ bar one. Although the relative error for these extremal
values increases with $N$, in practice however, these extremal values
will hardly ever occur, exactly because of the presence of dark
counts, as indicated by $a$. Therefore, given $a$ and $N$, we first
find the minimal value of the $g_i$ that can sensibly occur and then
calculate the relative error for that point. Since $\bm{G}$ is
distributed as a truncated multinomial one should take $g_i\ge
Na-2\sqrt{Na(1-a)}$. The relative error for points within these
boundaries is then less than $0.1\sigma$, independently of $N$.

We have compared the speed of three methods to calculate/approximate
the moments of $\bm{P}$. The calculations have been done in Matlab,
with the routines for the incomplete beta and incomplete gamma
function replaced by proprietary C implementations (available from
\cite{extra}). Method 1 is the saddle-point method combined with one
numerical integration (see appendix), method 2 is the saddle-point
method combined with analytical integration of a Taylor series
approximation (see appendix), and method 3 uses approximation
(\ref{eq:Jbybeta}). For $K=3$, $a=0.1$ and $\bm{\alpha}=[10,10,50]$,
method 1 took 142ms, method 2 10ms, and method 3 1.7ms, on an Intel
Core2 duo T7250 CPU running at 2GHz. Method 1 is the most accurate,
and method 3 the least.

\section{Dealing with Parameter Imprecision\label{sec:param}}
In the previous section we have assumed that the two main parameters
$\alpha$ and $\beta$ (dark count rate and attenuation factor) are
known exactly. In realistic situations, however, $\alpha$ and
$\beta$ are also of a statistical nature, for a variety of possible
reasons, including instability of the parameter (drift), imprecision
of the measurement of the parameter, or plain infeasibility of direct
measurement. The second best thing to an accurate value for a
parameter is then a statistical description in terms of a PDF or, at
the very least, in terms of its mean and central moments (variance,
and maybe even the skewness).

In this section we show how this statistical uncertainty about the
parameters can be included in the inference process. For simplicity of
the exposition, we will assume that only one parameter exhibits
imprecision. The general case follows easily.

Suppose, as usual, that we want to obtain an estimate of the random
variable $P$ and of its variance from measurements of $\bm{G}$, using
the likelihood function $L_{P|\bm{G},Y}(p|\bm{g},y)$, where $y$ is a
parameter that is described by a random variable $Y$, with given mean
$\mu$, variance $\sigma^2$ and possibly higher order moments.

We will assume that the PDF of $Y$ is close to normal, namely
continuous, single mode, small skewness and kurtosis close to the
normal value of 3. Almost all of the probability mass of $Y$ is then
contained in the interval $[\mu-3\sigma,\mu+3\sigma]$. PDFs of this
kind can be well approximated by a so-called Edgeworth expansion
\cite{hall,blinnikov}. A second order Edgeworth PDF is just the normal
PDF with the given mean and variance:
$$
f_{Y,2}(y)=\frac{1}{\sqrt{2\pi}\sigma}\,\exp[-(y-\mu)^2/2\sigma^2].
$$
A third order Edgeworth PDF adds another term, which contains the
skewness $\gamma$. For a standardised random variable (zero-mean and
unit variance) this PDF reads
$$
f_{Y,3}(y) = \phi(y) - \frac{\gamma}{6} \phi'''(y) =
[1-\gamma(3-y^2)y/6] \phi(y),
$$
where $\phi$ is the standardised normal PDF
$\phi(y)=\exp(-y^2/2)/\sqrt{2\pi}$.

Recall that if $Y$ were known perfectly, we would need to calculate
only the following:
\beas
\E[P] &=& \frac{\int_0^1 \md p\,\,p\,\,
L_{P|\bm{G}}(p|\bm{g},y)}{\int_0^1 \md p\,L_{P|\bm{G}}(p|\bm{g},y)},\\
\E[P^2] &=& \frac{\int_0^1 \md p\,\,p^2\,\,
L_{P|\bm{G}}(p|\bm{g},y)}{\int_0^1 \md p\,L_{P|\bm{G}}(p|\bm{g},y)},
\eeas
i.e.\ 3 integrals in total. Since, however, $Y$ enters as a nuisance
parameter, we must also integrate out $Y$, taking into account the PDF
of $Y$. Hence we need three \textit{double} integrals, which we would
like to avoid for efficiency reasons.

The method we will employ to simplify these calculations is to first
perform the integration over $p$ (analytically or numerically,
depending on what is possible), then approximate each such integral by
a polynomial of low degree (3 or 4) in $y$, (this is the idea behind 
the Newton-Cotes integration formulas) and finally perform the
integration over $Y$ analytically, with a low-order Edgeworth PDF
substituted for the PDF of $Y$.

To obtain a polynomial approximation we will use Lagrange
interpolation. Let $y_i$ be $m$ equidistant points within the interval
$[\mu-3\sigma,\mu+3\sigma]$ (with $m$ equal to 3 or 4), say
$y_i=i\delta$, with $i=-1,0,1$ or $i=-3/2,-1/2,1/2,3/2$. Then any
function $h(y)$ can be approximated by a polynomial $\hat{h}(y)$ given
by Lagrange's interpolation formula
$$
\hat{h}(y) = \sum_k h(y_k)\,\prod_{i, i\neq k} \frac{y-y_i}{y_k-y_i}.
$$

The integration over $Y$ can now be done analytically, provided we
choose a low-order Edgeworth PDF for $Y$. For $m=3$ (degree-3
interpolation) and choosing a normal PDF for $Y$ yields
\beas
&&\int \md y\,\,f_Y(y)\,\hat{h}(y)
\\&&
=\frac{\sigma^2}{2\delta^2} h(y_{-1})
+(1-\frac{\sigma^2}{\delta^2}) h(y_{0})
+\frac{\sigma^2}{2\delta^2} h(y_{1}).
\eeas
Hence, if we set $\delta=\sigma$, this formula simplifies to
\be
\int \md y\,\,f_Y(y)\,\hat{h}(y) =
[h(\mu-\sigma)+h(\mu+\sigma)]/2.
\ee
Hence, only two evaluations of $h$ are needed, i.e.\ two integrations
over $p$. As this has to be done for the numerator and denominator of
$\E[P]$ and of $\E[P^2]$, this gives a total of 6 integrations. For
example, the formula for $\mu_P$ becomes
$$
\E[P] = \frac{\int_0^1 \md p\,\,p\,\,L(p|\bm{g},\mu-\sigma)
+ \int_0^1 \md p\,\,p\,\,L(p|\bm{g},\mu+\sigma)}
{\int_0^1 \md p\,\,L(p|\bm{g},\mu-\sigma)
+ \int_0^1 \md p\,\,L(p|\bm{g},\mu+\sigma)}.
$$

For $m=4$ we can include the skewness $\gamma$ of $Y$ -- it cancels
out for $m=3$ -- by choosing a third-order Edgeworth PDF for $Y$.
When we put $\delta=2\sigma$, so that the whole $\pm3\sigma$ interval
is covered, we get in a similar way as before
\bea
&&
\int \md y\,\,f_Y(y)\,\hat{h}(y)
\nonumber\\&=&
-\frac{\gamma}{48} h(\mu-3\sigma)
+(\frac{1}{2}+\frac{\gamma}{16}) h(\mu-\sigma)\nonumber\\
&&+(\frac{1}{2}-\frac{\gamma}{16}) h(\mu+\sigma)
+\frac{\gamma}{48} h(\mu+3\sigma).
\eea
This now involves 4 evaluations of $h$, hence 4 integrals over $p$.

As a final remark, note that one can place bounds on the values of a
parameter from the measurement statistics. To illustrate this,
consider a run of $N$ 2-outcome pulsed experiments, with unknown dark
count rate, where the number $g$ of outcomes `1' is very low compared
to $N$. Intuition has it that the dark count rate must be small
accordingly. The likelihood function for $P$ is (see
Sec.~\ref{sec:singleB})
$$
L_P={N\choose g} [a+(1-2a)p]^g [a+(1-2a)(1-p)]^{N-g},
$$
with effective dark count rate $a$. Since $g$ is small, this places an
upper bound on the value of $a$. In effect, $a$ has to be described by
a random variable, and $L_P$ contains that random variable. By
integrating out $P$ from $L_P$, we obtain a distribution for $a$. The
exact result is that the PDF of $a$ is proportional to
\beas
f(a)&\propto& \int_0^1 \md p\,[a+(1-2a)p]^g [a+(1-2a)(1-p)]^{N-g} \\
&=& \frac{1}{1-2a}\int_a^{1-a} \md x\, x^g(1-x)^{N-g} \\
&=& \frac{B(a,1-a,g+1,N-g+1)}{1-2a}.
\eeas
Rather than using the exact result here, one notes that the integrand
of the second integral is proportional to the PDF of a beta
distribution and therefore $f(a)$ is essentially the cumulative distribution function (CDF)
of the complementary
beta distribution, a function decreasing with $a$. The PDF has mean value
$\mu=(g+1)/(N+2)$ and variance
$\sigma^2=(g+1)(N+1-g)/(N+2)^2(N+3)$. Thus, $f(a)$ will be significant
only for values of $a$ below $\mu+3\sigma$. For small $g$ and large
$N$, we therefore get the promised upper bound on $a$:
\be
a\le(g+1+3\sqrt{g+1})/N.
\ee

\section{Poissonian case\label{sec:poisson}}

In Sec.~\ref{sec:pulses} we have treated a class of tomography
experiments based on single-photon optical pulses, where the
statistics of the recorded photon counts is governed by the
binomial/multinomial distribution. In this section we treat continuous
wave (CW) experiments. Here, the input laser beam is turned on for a
fixed time $T$. The detectors are still operating in Geiger mode, and
the intensity of the laser beam is such that individual photons can
still be discerned. Photon counts are recorded during that same time
interval $T$. The statistics are now governed by the Poisson
distribution.

Note that the Poisson distribution is the limiting case of the
binomial distribution for the number of runs $N$ going to infinity,
while the total duration $T$ and the photon rate (average number of
photons expected during $T$) are kept constant. Therefore, in
principle, there should be no essential difference between the
statistics of this kind of experiment and those of the single-photon
experiments. However, in CW experiments, the intensity of the laser
beam enters as a parameter, requiring determination. While this
determination is possible by performing independent measurements, a
less time-consuming approach is to use the actual measurements one is
interested in. This approach will be described in this section.

We will assume again that the dark count rate $\alpha$ is known exactly. The 
detector attenuation factor $\beta$ will not show up explicitly as it is assumed to be absorbed
into the (unknown) laser beam intensity.
\subsection{Statistical Model}
We consider a CW experiment consisting of $K$ runs of equal time
duration $T$, and constant but unknown laser intensity. In each run a
different 2-outcome POVM $\{\Pi^{(i)},\id-\Pi^{(i)}\}$ is applied, but
only the counts $g_i$ corresponding to $\Pi^{(i)}$ are recorded, as
was the case in Sec.~\ref{sec:singleA}. We assume that $\sum_i
\Pi^{(i)}=b\id$. The general case, in which $\sum_i \Pi^{(i)}$ is not
a multiple of $\id$, has been treated (without dark counts) in
Ref.~\cite{kalman1}. The purpose of this section is only to show how dark
counts can be added to the statistical model. Non-unit detector
efficiency has already been incorporated in the treatment of
Ref.~\cite{kalman1} implicitly, by absorbing $\eta$ in the beam
intensity $A$.

As stated in Sec.~\ref{sec:model}, for Poissonian input and
background fields, the counts are Poissonian too, with mean value
$\mu=\alpha+\eta\nu$, where $\alpha$ is the dark count rate and $\nu$
the input photon rate. For beam intensity $A$, and POVM element
$\Pi^{(i)}$, we have $\nu =Ap_i$, thus $\mu_i=\alpha+\eta A
p_i$. Henceforth, we absorb $\eta$ into $A$, thus $\mu_i=\alpha+A
p_i$. In addition, since $\sum_i \Pi^{(i)}=b\id$, we have $\sum_i p_i=b$.

As the counts $g_i$ are independent, and each is Poissonian with mean
$\mu_i$, the PDF of the sequence of counts
$\bm{G}=(G_1,G_2,\ldots,G_K)$ is given by
\beas
f_{\bm{G}}(\bm{g}) = \prod_{i=1}^K e^{-\mu_i}
\frac{\mu_i^{g_i}}{g_i!}
\propto
e^{-Ab}\prod_{i=1}^K (\alpha+A p_i)^{g_i},
\eeas
where factors have been left out that are independent of $p_i$ and
$A$. In order to formally turn the quantities $\alpha+A p_i$ into a
probability distribution, we divide by their sum
$\sum_{i=1}^K(\alpha+A p_i) = K\alpha+Ab$, and define
\bea
\frac{\alpha+A p_i}{K\alpha+Ab} &=& y+(1-Ky)p_i/b, \\
y&:=& \alpha/x, \\
x&:=& K\alpha+Ab.
\eea
Then the PDF of $\bm{G}$ is proportional to
\be
f_{\bm{G}}(\bm{g})
\propto \frac{e^{-x}x^N}{\Gamma(N+1,K\alpha)} \,\, \prod_{i=1}^K
(y+(1-Ky) p_i/b)^{g_i},\label{eq:fff}
\ee
with $N:=\sum_i g_i$. The factor $1/\Gamma(N+1,K\alpha)$ has been
included to normalise the factor $e^{-x}x^N$ over the interval $x\ge
K\alpha$. The first factor is, indeed, the PDF of a truncated gamma
distribution.

The second factor is essentially the PDF for the single-photon case,
with $y$ assuming the role of the effective dark count rate. The main
difference is that $y$ is now a random variable. Indeed, as the
variable $A$ is an unknown, so are $x$ and $y$. In Bayesian
terminology, $A$ is a nuisance parameter, and the standard Bayesian
treatment is to integrate it out. That is, $f_{\bm{G}}(\bm{g})$ is
multiplied by a suitable prior for $A$, and is then integrated over
$A\in[0,\infty]$. The problem with this approach is that the integral
cannot be carried out analytically.

In what follows, we approximate the integral, based on the assumption
that the number of total counts $N=\sum_i g_i$ should be much larger
than the expected total number of dark counts $K\alpha$, i.e.\ that
the signal-to-noise ratio of the experimental data is large
enough. This assumption is very reasonable given that one actually
wants to obtain useful information from the data.

The main benefit of this assumption is that the truncation of $x$ can
be disregarded. Indeed, as has been noted in Sec.~\ref{sec:prelim},
the normalisation factor $\Gamma(N+1,K\alpha)$ is well approximated by
$\Gamma(N+1)$ when $N\ge 1+K\alpha+3\sqrt{K\alpha}$, and similar
statements hold regarding the moments of the distribution. Thus the
PDF of $\bm{G}$ is proportional to
\be
f_{\bm{G}}(\bm{g})
\propto \frac{e^{-x}x^N}{N!} \,\, \prod_{i=1}^K
[y+(1-Ky) p_i/b]^{g_i},\label{eq:fff2}
\ee
where we now allow the random variable $X$ to assume all values down
to 0. The upshot is that to very good approximation, $X$ has a gamma
distribution with mean (and variance) $N+1$. The integral of
$f_{\bm{G}}(\bm{g})$ over $x$ is thus a convolution of
$\cL(\bm{g}):=\prod_{i=1}^K [y+(1-Ky) p_i/b]^{g_i}$, which depends on
$x$ via $y=\alpha/x$, with the gamma PDF of $X$. Note also the
resemblance of Eq.~(\ref{eq:fff2}) to the corresponding
Eq.~(\ref{eq:fgpmulti}) for the $K$-detector single photon case, which
is not all too surprising.

A short calculation using the properties of $X$ reveals that the
variable $Y=\alpha/X$ has mean value $\mu_Y = \alpha/N$ and variance
$\sigma_Y^2=\alpha^2/N^2(N-1)$. As the PDF of $Y$ shows small but
noticeable deviations from a normal distribution, we also need the
skewness of $Y$, which turns out to be $\gamma_Y=4\sqrt{N-1}/(N-2)$.
Recall that the skewness is defined as the third central moment of $Y$
divided by the third power of $\sigma_Y$; for this distribution the
skewness is roughly equal to two times Pearson's mode skewness, and
can therefore be interpreted as how much the mean differs from the
mode, expressed in halves of a standard deviation. For this
distribution the mode of $Y$ is $\alpha/(N+2)$.

\subsection{Statistical Inference}
We can now invoke the methods of Secs.~\ref{sec:mtnK} and
\ref{sec:param} to perform the statistical inversion of $\cL(\bm{g})$
with $Y$ as an imprecise parameter with the moments just mentioned, which depend
on the dark count rate $\alpha$ (assumed to be known here) and on $N=\sum_i g_i$. 
As regards the additional factor $1/b$ in Eq.~(\ref{eq:fff2}),
this can be taken into account by multiplying the obtained mean of
$\bm{P}$, $\E[P_i]$, by $b$ and the second order moments about the
origin, $\E[P_iP_j]$, by $b^2$.

Finally, we can also treat the case where the POVM elements do not add
up to a multiple of the identity, i.e.\ when the assumption $\sum_i
\Pi^{(i)}=b\id$ is not satisfied. This could occur because of
inaccuracies in the implementations of the POVM elements, or simply
because of the choice of elements -- before Ref.~\cite{kalman1} it was
not known that failure to meet the condition $\sum_i \Pi^{(i)}=b\id$
had a severely negative impact on the ease with which statistical
inferences could be made. The consequence is that the probabilities
$p_i$ do not add up to a constant. Their sum $p_0:=\sum_{i=1}^K p_i$
is now a random variable, too, and has to be treated as an additional
nuisance parameter. This case has been treated, for the case without
dark counts, in Ref.~\cite{kalman1}, Sec.~3.2.5, under the assumption
that the deviation of $\sum_i \Pi^{(i)}$ from a scalar matrix is
small. For larger deviations no accurate methods are known to us other
than Monte-Carlo methods.

The formulas obtained in Ref.~\cite{kalman1} carry over easily to the
case with dark counts, because $b$ simply enters as a factor in the
formulas for the moments of $\bm{P}$. Let $M$ and $m$ be the largest
and smallest eigenvalue of $\sum_i\Pi^{(i)}$. The multiplication
factors for $\E[P_i]$ and $\E[P_iP_j]$ (the moments about the origin)
are now, instead of $b$ and $b^2$, $M\phi_1$ and $M\phi_2$,
respectively, with
\bea
\phi_1 &=& \frac{K}{K+1}\,\frac{1-(m/M)^{K+1}}{1-(m/M)^K}, \\
\phi_2 &=& \frac{K}{K+2}\,\frac{1-(m/M)^{K+2}}{1-(m/M)^K}.
\eea
\section{Conclusion\label{sec:conclusion}}
In this paper, we have studied the statistical properties of photon
detection using imperfect detectors,  exhibiting dark counts and less
than unit detection efficiency, in the context of implementations of
general $K$-element POVMs. We have derived a Bayesian inference
procedure for obtaining distributions over outcome probabilities  from
detection frequencies in a variety of setups. We also obtained
formulas and/or algorithms for efficiently calculating the first and
second order moments of these distributions, effectively obtaining
estimates and corresponding error bars for the outcome probabilities.

For experiments using single-photon laser pulses we have considered
$K$-element POVMs constructed with $K$ detectors (with special
emphasis on the case $K=2$). We found that by far the easiest
inference procedure occurred when only taking single-detection events
into account (i.e.\ only counting events where just one out of $K$
detector clicked). In that case, the outcome probabilities $\bm{p}$
are drawn from a truncated Dirichlet distribution
$\propto \prod_{i=1}^K [a+(1-Ka) p_i]^{g_i}$ where $g_i$ are the
detection frequencies and $a$ is an effective dark count
rate, which can be calculated from the actual dark count rate and the
detection efficiency. For $K=2$ the moments of this truncated
Dirichlet can be calculated extremely rapidly using incomplete beta
functions. For larger $K$ we have devised a number of numerical
algorithms for doing so, offering the user a trade-off between
accuracy and speed. For $K=2$ we also considered a setup with just a
single detector, and found slightly different formulas for the
distribution and its moments.

While in the above one needs to supply values for dark count rate and
detector efficiency, we have also devised a method for dealing with
the case when these parameters are not accurately known. This method
is particularly useful to deal with the final setup we have considered,
namely when the experiments are done with continuous wave laser
beams. In that case, the detection statistics is Poissonian and the
inferred outcome probabilities are again drawn from a truncated
Dirichlet, but now with the  effective dark count rate being a random
variate itself, due to the inaccurately unknown laser beam intensity.

Finally, we also briefly considered how one can obtain an upper bound
on the effective dark count rate, from the value of the minimal
frequency of an outcome in any given run (or in a combination of
runs).
\acknowledgments
KA thanks Tobias Osborne for discussions when this work was still in 
its infantile stage. 
SS thanks the UK Engineering and Physical Sciences Research Council
(EPSRC) for support.

\appendix
\section{Integrals of truncated Dirichlet distributions\label{app:A}}
In order to calculate the moments of the truncated Dirichlet
distribution, one must be able to accurately calculate the
distribution's normalisation integrals. In this Appendix, we describe
an approximation method due to Butler and Sutton \cite{walrus}.

Let $\bm{X}\sim\mbox{Dir}(\bm{\alpha})$ be a Dirichlet distributed
$K$-dimensional random variable, with parameters
$\bm{\alpha}=(\alpha_1,\ldots,\alpha_K)$. This assumes that $X_i\ge0$
and $\sum_{i=1}^K X_i=1$ hold. We will use the common notation
$\alpha_0=\sum_i\alpha_i$.

Let us now truncate $\bm{X}$, by imposing the condition $X_i\ge a$,
where $0\le a\le 1/K$. The goal is to calculate the new integration
constant given by the probability $\pr(\bm{X}\ge a)$. We will denote
this probability integral by $J$:
\be
J(\bm{\alpha};a) =
\int_{\genfrac{}{}{0pt}{}{x_i\ge a}{\sum_i x_i=1}}
d\bm{x} \,\, \Gamma(\alpha_0)\,\,\prod_{i=1}^K
\frac{x_i^{\alpha_i-1}}{\Gamma(\alpha_i)}.
\ee
Note that for $K=2$, this integral is given by the regularised
incomplete beta function $I_{a,1-a}(a_1,a_2)$.

The method proposed by Butler and Sutton consists of two basic
ideas. The first idea is to use a conditional characterisation of
$\bm{X}$. Namely, one defines $K$ new, \textit{independent} random
variables $Z_i$ such that $\bm{X}$ and $\bm{Z}|\sum_i Z_i=1$ have the
same distribution. It is known that one obtains the required Dirichlet
distribution if $Z_i$ has a gamma distribution, $Z_i\sim
\mbox{Gamma}(\alpha_i,1)$. For the purposes of the method, the value
of the scale parameter $\theta$ does not matter, and we set
$\theta=1$. The PDF is therefore given by
$$
f_{Z_i}(z) = \frac{z^{\alpha_i-1}e^{-z}}{\Gamma(\alpha_i)}.
$$

Now the required probability $\pr(\bm{X}\ge a)$ can be expressed,
using Bayes' rule, as
\bea
&& \pr(\bm{X}\ge a) = \pr(\bm{Z}\ge a|\sum_i Z_i=1) \nonumber \\
&=& \pr(\sum_i Z_i=1|\bm{Z}\ge a)
\prod_i \pr(Z_i\ge a)
\frac{1}{\pr(\sum_i Z_i=1)}.\nonumber \\ \label{eq:A1}
\eea

The factors $\pr(Z_i\ge a)$ are easily calculated in terms of the CDF
of the gamma distribution, giving
\be
\pr(Z_i\ge a) = Q(\alpha_i,a),
\ee
with $Q(\alpha_i,a)$ the regularised incomplete gamma function.

Since the $Z_i$ are independently gamma-distributed, $Z_i\sim
\mbox{Gamma}(\alpha_i,1)$, their sum is also gamma-distributed:
$\sum_i Z_i\sim \mbox{Gamma}(\alpha_0,1)$. The factor $\pr(\sum_i
Z_i=1)$ is therefore given by the value of the PDF of
$\mbox{Gamma}(\alpha_0,1)$ in 1, which gives:
\be
1/\pr\left(\sum_i Z_i=1\right) = e\,\Gamma(\alpha_0).
\ee

The first factor in Eq.~(\ref{eq:A1}), the truncated PDF $\pr(\sum_i
Z_i=1|\bm{Z}\ge a)$, is the hardest to calculate, because it is a
multi-dimensional integral, and the second idea in Butler and Sutton's
method is to convert it to an inverse Laplace integral of a univariate
function, and then approximate the latter integral using a
saddle-point method, as first proposed by Daniels \cite{daniels}.

The method starts from the moment generating function (MGF) of the
truncated random variable $T=\sum_i Z_i|Z\ge a$, defined as
$M_T(s)=\E_T[e^{st}]$. Since the $Z_i$ are independent, we have
\be
M_T(s) = \prod_i \E_{T_i}[e^{st}],
\ee
where $T_i:=Z_i|Z_i\ge a$.
A simple calculation gives
\bea
\E_{T_i}[e^{st}] &=& \frac{\int_a^\infty dt\,\, e^{st}t^{\alpha_i-1}e^{-t}}
{\int_a^\infty dt\,\, t^{\alpha_i-1}e^{-t}} \nonumber \\
&=& (1-s)^{-\alpha_i} \frac{Q(\alpha_i,(1-s)a)}{Q(\alpha_i,a)},
\eea
which is valid for $\Re s<1$ (and we do need complex $s$).
The denominators cancel with the factors $\pr(Z_i\ge a) = Q(\alpha_i,a)$.

Since the MGF $M_T(-s)$ is the two-sided Laplace transform of the PDF,
the PDF can be recovered from the MGF by an inverse Laplace transform:
$$
f_T(t) = \frac{1}{2\pi i}\int_{\gamma-i\infty}^{\gamma+i\infty} M_T(s)
e^{-st} ds,
$$
where, in our case, we only need to evaluate the PDF at the point
$t=1$. By expressing the MGF as the exponential of the cumulant
generating function (CGF) $K_T(s):=\log M_T(s)$, the path of
integration can be brought in a form that readily invites the
saddle-point method for its approximate evaluation:
\be
f_T(t) = \frac{1}{2\pi i}\int_{\gamma-i\infty}^{\gamma+i\infty}
e^{K_T(s)-st} ds. \label{eq:sad1}
\ee
The path of integration is hereby chosen to pass through a
saddle-point of the integrand, in such a way that the integrand is
negligible outside its immediate neighbourhood. Daniels shows that in
this case the path should be a straight line parallel to the imaginary
axis and passing through the saddle-point $\hat{s}$, which is that
value of $s$ for which the derivative of $K_T(s)-st$ w.r.t.\ $s$
vanishes:
\be
K'_T(\hat{s})=t.
\ee
Daniels showed that, under very general conditions, $\hat{s}$ is real.
Hence, in Eq.~(\ref{eq:sad1}), one takes $\gamma=\hat{s}$, and the
path of integration is along points $s=\hat{s}+iy$.

An explicit formula for $K'_T(s)$ is
\be
K'_T(s) = \frac{a}{u} \left[\alpha_0 +\sum_i g(\alpha_i,u)\right],
\ee
with $u=a(1-s)$ and $g(\alpha,u) = e^{-u} u^{\alpha_i}/\Gamma(\alpha_i,u)$.
One shows that $g(\alpha,u)$ is roughly approximated by
$\max[0,u-(\alpha-1)]$; moreover,
$g(\alpha,u)\ge \max[0,u-(\alpha-1)]$.  An approximate value of
$\hat{s}=1-\hat{u}/a$ is thus given by the solution of
\be
\frac{u}{a} = \alpha_0+\sum_i \max[0,u-(\alpha_i-1)].
\ee
As the right-hand side is a piecewise linear function of $u$, the
solution of this equation is easily found. This approximate solution
can then be used as a starting value for numerically solving the exact
equation
$$
\frac{u}{a} = \alpha_0+\sum_i g(\alpha_i,u).
$$

Once the optimal value $\hat{s}$ has been obtained, one can go about
performing the integration in Eq.~(\ref{eq:sad1}), i.e.\ of
\be
f_T(1) = \frac{1}{\pi}\int_0^{\infty}
\Re[M_T(\hat{s}+iy)e^{-(\hat{s}+iy)}] dy, \label{eq:sad2}
\ee
where we have exploited the fact that the real part of the integrand is
even in $y$. To obtain the highest accuracy, the integration has to be
done using a numerical quadrature (e.g.\ using Matlab's built-in
{\tt quadl} routine). The upper integration limit can be replaced by a
finite value, equal to a fixed number times the approximate width of
the function graph, which is roughly $1/\sqrt{K''_T(\hat{s})}$, where
$$
K''_T(s) = \sum_i \frac{\alpha_i}{(1-s)^2\Gamma(\alpha_i,a(1-s))^2}.
$$

If speed is at a premium, while somewhat less precision is acceptable,
one can use a finite-term Taylor expansion of $K_T(s)-s$, and integrate
each of the resulting terms analytically. The saddle-point
approximation is obtained by writing $K_T(s)-s$ as a Taylor series
around $s=\hat{s}$:
$$
K_T(s)-s = K_T(\hat{s})-\hat{s}+\sum_{j=2}^\infty \frac{1}{j!}
K^{(j)}_T(\hat{s}) (iy)^j,
$$
and expanding the integrand as
\beas
e^{K_T(s)-s} &=& e^{K_T(\hat{s})-\hat{s}} \,\,e^{-K''_T y^2/2}
\exp\left[\sum_{j=3}^\infty \frac{1}{j!} K^{(j)}_T (iy)^j\right] \\
&=& e^{K_T(\hat{s})-\hat{s}} \,\,e^{-K''_T y^2/2} \\
&&\times\Bigg\{1-i \frac{K^{(3)}_T}{6}y^3+\frac{K^{(4)}_T}{24}y^4
+i\frac{K^{(5)}_T}{120}y^5 \\
&& +\left[-\frac{(K^{(3)}_T)^2}{72}-\frac{K^{(6)}_T}{720}\right]y^6
+\ldots\Bigg\}.
\eeas
with each of the derivatives of $K_T$ evaluated in $\hat{s}$.

\begin{figure}[ht]
\includegraphics[width=8cm]{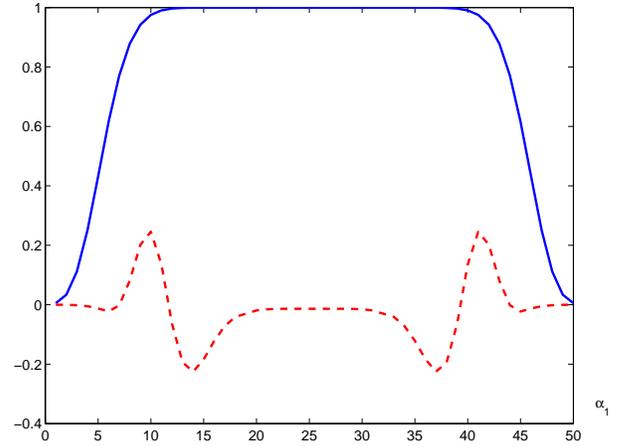}
\caption{
\label{fig:trunc}
Plot of $J(\alpha_1,\alpha_2;0.1)$ as calculated using the second-order
saddle-point method (blue, solid curve),
and the absolute error, in units of $10^{-4}$ (red, dashed curve),
as compared to the exact result $I_{0.1,0.9}(\alpha_1,\alpha_2)$.
The sum $\alpha_0=\alpha_1+\alpha_2$ is held constant at a value of 50.
The maximal absolute error here is $2.4571\times 10^{-5}$
and the maximal relative error is $2.5189\times 10^{-5}$.
}
\end{figure}
Upon performing the integral $\int_{-\infty}^{+\infty} dy$ the terms
with odd powers of $y$ vanish. After substituting $K''_T y^2/2=v^2$,
and using
$$
\int_{-\infty}^{+\infty} e^{-v^2}v^{2k}dv = \Gamma(k+1/2),
$$
with $\Gamma(1/2)=\sqrt{\pi}$, $\Gamma(2+1/2)=3\sqrt{\pi}/4$ and
$\Gamma(3+1/2)=15\sqrt{\pi}/8$, the even powers yield
\be
f_T(1) = \frac{e^{K_T(\hat{s})-\hat{s}}}{\sqrt{2\pi K''_T}}
\left(1+\frac{1}{8}\,\frac{K^{(4)}_T}{(K''_T)^2}-\frac{5}{24}\,
\frac{(K^{(3)}_T)^2}{(K''_T)^3}+\ldots\right),
\ee
(note that in the corresponding formula (7) in Ref.~\cite{walrus} a
minus sign is missing).

In Fig.~\ref{fig:trunc} we give an example of the 2-dimensional
integral $J(\alpha_1,\alpha_2;a)$ calculated using this method and
compare it to the exact result, which for $K=2$ is known to be the
regularised incomplete beta function
$I_{a,1-a}(\alpha_1,\alpha_2)$. The Matlab routines used to perform
these calculations are available from \cite{extra}. 
\section{Mathematical compendium} 
\label{sec:prelim}
In this appendix we gather a few mathematical preliminaries that are necessary to
understand the statistical models developed in
Secs.~\ref{sec:model}--\ref{sec:poisson}.

\subsection{Special functions}
We start by collecting some important results on special functions and
their implementations in various computer algebra software.

\subsubsection{Gamma function}
The \textbf{gamma function} $\Gamma(\alpha)$ is defined as the
integral
\be
\Gamma(\alpha) = \int_0^\infty \md t\,t^{\alpha-1} e^{-t}
\ee
with $\Gamma(k)=(k-1)!$ for integer arguments.
Since for large values of its argument, the gamma function becomes
extremely large, numerical packages usually contain implementations of
the natural logarithm of the gamma function too ({\tt gammaln} in
Matlab, and {\tt LogGamma} in Mathematica). We will need this as well.

The gamma integral leads to two incomplete integrals,
the \textbf{lower incomplete gamma function} $\gamma(\alpha,x)$
and the \textbf{upper incomplete gamma function} $\Gamma(\alpha,x)$:
\bea
\gamma(\alpha,x) &=& \int_0^x \md t\,t^{\alpha-1} e^{-t}, \\
\Gamma(\alpha,x) &=& \int_x^\infty \md t\,t^{\alpha-1} e^{-t}.
\eea
Obviously, one has $\gamma(\alpha,x)+\Gamma(\alpha,x)=\Gamma(\alpha)$.
By dividing these incomplete gamma functions by the corresponding
complete gamma, one obtains the
\textbf{regularised incomplete gamma functions}:
\bea
P(\alpha,x) &=& \gamma(\alpha,x)/\Gamma(\alpha), \\
Q(\alpha,x) &=& \Gamma(\alpha,x)/\Gamma(\alpha),
\eea
with $P+Q=1$.

In Mathematica, {\tt Gamma[$\alpha$,x]} is the upper incomplete gamma
function $\Gamma(\alpha,x)$, while {\tt Gamma[$\alpha$,$x_0$,$x_1$]}
is the generalized incomplete gamma function, so that
$\gamma(\alpha,x)=$ {\tt Gamma[$\alpha$,0,x]}.
The regularised incomplete gamma functions are implemented as
$Q(\alpha,x)=$ {\tt GammaRegularized[$\alpha$,x]} and
$P(\alpha,x)=$ {\tt GammaRegularized[$\alpha$,0,x]}.

In Matlab, $P(\alpha,x)$ has been implemented as
{\tt gammainc(x,$\alpha$)} (note the reversal of the arguments). Except in older versions,
$Q$ has been implemented too, as
{\tt gammainc(x,$\alpha$,'upper')}.

The two basic expansions that are used in these calculations
are the series expansion (see, e.g. Ref.~\cite{as}, formula 6.5.29)
$$
P(\alpha,x) = e^{-x} \, \sum_{k=0}^\infty
\frac{x^{\alpha+k}}{\Gamma(\alpha+k+1)},
$$
for $x<\alpha+1$,
and the continued fraction expansion (see, e.g. Ref.~\cite{as},
formula 6.5.31)
$$
Q(\alpha,x)=\frac{e^{-x}x^\alpha}{\Gamma(\alpha)}\left(\frac{1}{x+}
\,\,\frac{1-\alpha}{1+}\,\,
\frac{1}{x+}\,\,\frac{2-\alpha}{1+}\,\,\frac{2}{x+}\ldots\right),
$$
for $x\ge \alpha+1$.
Here we used the typographical notation for continued fractions:
$\left(\frac{a}{b+}c\right) = \frac{a}{b+c}$, where $c$ stands for everything that follows.
For the other regimes one can use the formula $P+Q=1$.
If high accuracy is needed for extremely small values of $P$ or $Q$,
one should calculate the logarithm.
\subsubsection{Beta function}
The \textbf{beta function} $B(a,b)$, a generalization of the gamma
function,  is defined as
\be
B(a,b) = \int_0^1 \md t\,t^{a-1}(1-t)^{b-1}.
\ee
It is related to the gamma function via
\be
B(a,b) = \frac{\Gamma(a)\Gamma(b)}{\Gamma(a+b)}.
\ee
This leads to the relation
\be
B(a+1,b)/B(a,b) = a/(a+b),
\ee

For integer arguments, one sees that $B(a,b)$ is related to the
binomial coefficient as
$$
B(a,b)=\frac{(a-1)!(b-1)!}{(a+b-1)!} = \frac{a+b}{ab{a+b\choose a}}.
$$
Since, again, the natural logarithm of the beta function is usually
implemented directly [in Matlab: {\tt betaln(a,b)}],
this formula allows evaluation of the binomial coefficients for larger
values of the arguments than allowed by direct calculation.

Just as in the case of the gamma function, replacing the integration
limits yields the \textbf{incomplete beta function} $B(x,a,b)$
and the \textbf{generalised incomplete beta function} $B(x_0,x_1,a,b)$
\bea
B(x,a,b) &=& \int_{0}^{x} \md x\,x^{a-1}(1-x)^{b-1}\\
B(x_0,x_1,a,b) &=& \int_{x_0}^{x_1} \md x\,x^{a-1}(1-x)^{b-1}.
\eea
Dividing by the complete beta function also gives the
\textbf{regularised incomplete beta function} and the
\textbf{generalised regularised incomplete beta function}
\bea
I_{x}(a,b) &=& B(x,a,b)/B(a,b), \label{eq:regincbeta}\\
I_{x_0,x_1}(a,b) &=& B(x_0,x_1,a,b)/B(a,b).
\eea

In Matlab, only $I_{x}(a,b)=I_{0,x}(a,b)$ and
$I_{x,1}(a,b)=1-I_x(a,b)$ are implemented, as {\tt betainc(x,a,b)}
and {\tt betainc(x,a,b,'upper')}, the latter only in more recent
versions, while in Mathematica all four functions exist, under the
names {\tt Beta[x,a,b]}, {\tt Beta[x0,x1,a,b]},
{\tt BetaRegularized[x,a,b]} and {\tt BetaRegularized[x0,x1,a,b]}.
Just as for the incomplete gamma functions one may need a logarithmic
version of $I_x$ to cover cases with extremely small function values.

Calculations are based on the continued fraction expansion
of $I_x$, which is valid for $x$ smaller than $(a-1)/(a+b-2)$
(see, e.g. Ref.~\cite{as}, formula 26.5.8):
\be
I_x(a,b) \approx \frac{x^a(1-x)^b}{a B(a,b)}
\left(\frac{1}{1+}\,\,\frac{d_1}{1+}\,\,\frac{d_2}{1+}\ldots\right)
\ee
with
\beas
d_{2m+1} &=& -\frac{(a+m)(a+b+m)}{(a+2m)(a+2m+1)}\,\,x, \\
d_{2m} &=& \frac{m(b-m)}{(a+2m-1)(a+2m)}\,\,x.
\eeas
For larger $x$, one uses the relation $I_{1-x}(b,a)=1-I_x(a,b)$, where
the left hand side is numerically more accurate for small function
values. In case the continued fraction expansion fails, one can still
use certain approximations (see, e.g. Ref.~\cite{as}, formulas 26.5.20
and 21).

\subsection{Poisson, Gamma, Beta and Dirichlet Distributions
\label{sec:dirichlet}} 
The probability distribution function (PDF) of a discrete random
variable $K$ that is distributed according to the
\textbf{Poisson distribution}, $K\sim\Pe(\lambda)$, is
\be
f_K(k)=\frac{\lambda^k e^{-\lambda}}{k!}.
\ee
Its mean and variance are both equal to $\lambda$.

We also recall a number of basic facts about several continuous
distributions \cite{jkb94,kbj}. The \textbf{gamma distribution} is
directly related to the gamma function. The PDF of a random variable
$X$ that is distributed according to the gamma distribution
$X\sim\mbox{Gamma}(\alpha,\theta)$, with $\alpha$ the shape parameter
and $\theta$ the scale parameter, is given by
$$
f_X(x) = \frac{e^{-x/\theta}x^{\alpha-1}}{\theta^\alpha\Gamma(\alpha)}.
$$
We will not need the extra freedom offered by $\theta$, and we will
always put $\theta=1$, giving
\be
f_X(x) = \frac{e^{-x}x^{\alpha-1}}{\Gamma(\alpha)}.
\ee

For $x=\lambda$ and $\alpha=k+1$, this PDF looks formally the same as
the Poisson PDF. However, in the latter $K$ is the random variable,
rather than $X$. In effect, the gamma distribution and Poisson
distribution are each other's conjugate.

The cumulative distribution function (CDF) of $X$ is the regularised
lower incomplete gamma function $P$:
\be
\pr(X\ge x) = P(\alpha,x),
\ee
and its moments are given by
\be
\mu_X = \sigma_X^2 = \alpha.
\ee
For not too small values of $\alpha$, the bulk of the probability mass
of the gamma distribution is roughly contained within the interval
$[\mu-3\sigma,\mu+3\sigma]=[\alpha-3\sqrt{\alpha},\alpha+3\sqrt{\alpha}]$.
This explains why $P(\alpha,x)$ is very close to 0 for
$x\le\alpha-3\sqrt{\alpha}$ and very close to 1 for (roughly)
$x\ge\alpha+3\sqrt{\alpha}$. A more accurate statement is that for
$x\ge \alpha+2.8+3.09\sqrt{\alpha}$, or $\alpha\le
x+1.9-3.09\sqrt{x-0.41}$, $P(\alpha,x)\ge 0.999$.

The \textbf{Dirichlet distribution} is the higher-dimensional
generalisation of the beta distribution. The importance of this
distribution stems from the fact that it is the conjugate distribution
of the multinomial distribution: if $\bm{F}\sim \mbox{Mtn}(N,\bm{p})$
is the distribution of $\bm{F}$ conditional on $\bm{P}=\bm{p}$, then
using Bayesian inversion (starting with a uniform prior for $\bm{P}$)
$\bm{P}$ conditional on $\bm{F}=\bm{f}$ is Dirichlet distributed with
parameter $\bm{f}$. Formally, the two distributions only differ by
their normalisation. The multinomial distribution is normalised by
summing over all integer non-negative $\bm{f}$ summing up to $N$,
while the Dirichlet distribution is normalised by integrating over the
simplex of non-negative $\bm{p}$ summing to 1.


The general form of the PDF of a $d$-dimensional Dirichlet
distribution with parameters $\alpha_i$ is (see, e.g. Ref.~\cite{kbj},
Chapter 49)
$$
f_{\bm{P}}(\bm{p}) = \Gamma(\alpha_0) \prod_{i=1}^d
\frac{p_i^{\alpha_i-1}}{\Gamma(\alpha_i)},
$$
where $\alpha_0$ is defined as
\be
\alpha_0 := \sum_{i=1}^d \alpha_i.
\ee
The range of $\bm{P}$ is the simplex $p_i\ge0, \sum p_i=1$.

The mean values of the Dirichlet distribution are
\be
\mu_i = \frac{\alpha_i}{\alpha_0} ,\label{eq:dirichletmean}
\ee
and the elements of its covariance matrix are
\be
\sigma_{ij}^2 = \left\{
\begin{array}{ll}
\frac{\alpha_i(\alpha_0-\alpha_i)}{\alpha_0^2(\alpha_0+1)},&i=j \\[2mm]
\frac{-\alpha_i\alpha_j}{\alpha_0^2(\alpha_0+1)},&i\neq j
\end{array}
\right..\label{eq:dirichletcov}
\ee

The \textbf{beta distribution} is the special case of a Dirichlet
distribution with $d=2$. The normalisation factor is then the beta
function $B(\alpha_1,\alpha_2)$, from which the distribution got its
name.

\end{document}